\documentclass[aip,amsmath,amssymb,prereprint,superscriptaddress]{revtex4-1}
\usepackage{graphicx}
\usepackage{amsmath}
\usepackage{dcolumn}
\usepackage{bm}
\usepackage{bbold}
\usepackage{color}

\begin{document}


\title{Extremely Low Loss Phonon-Trapping Cryogenic Acoustic Cavities for Future Physical Experiments}

\author{Serge Galliou}
\affiliation{Department of Time and Frequency, FEMTO-ST Institute, ENSMM, 26 Chemin de l'\'{E}pitaphe, 25000, Besan\c{c}on, France}

\author{Maxim Goryachev}
\affiliation{ARC Centre of Excellence for Engineered Quantum Systems, University of Western Australia, 35 Stirling Highway, Crawley WA 6009, Australia}

\author{Roger Bourquin}
\affiliation{Department of Time and Frequency, FEMTO-ST Institute, ENSMM, 26 Chemin de l'\'{E}pitaphe, 25000, Besan\c{c}on, France}

\author{Philippe Abb\'{e}}
\affiliation{Department of Time and Frequency, FEMTO-ST Institute, ENSMM, 26 Chemin de l'\'{E}pitaphe, 25000, Besan\c{c}on, France}

\author{Jean Pierre Aubry}
\affiliation{Oscilloquartz SA, Br\'{e}vards 16, 2002 Neuch\^{a}tel, Switzerland, {now self-employed consultant (jeanpierre.aubry@net2000.ch)}}

\author{Michael E. Tobar}
\email{michael.tobar@uwa.edu.au}
\affiliation{ARC Centre of Excellence for Engineered Quantum Systems, University of Western Australia, 35 Stirling Highway, Crawley WA 6009, Australia}

\date{\today}


\begin{abstract}
Low loss Bulk Acoustic Wave devices are considered from the point of view of the solid state approach as phonon-confining cavities. 
We demonstrate effective design of such acoustic cavities with phonon-trapping techniques exhibiting extremely high quality factors for trapped longitudinally-polarized phonons of various wavelengths. Quality factors of observed modes exceed 1 billion, with a maximum $Q$-factor of 8 billion and $Q\times f$ product of $1.6\cdot10^{18}$ at liquid helium temperatures. Such high sensitivities allow analysis of intrinsic material losses in resonant phonon systems. Various mechanisms of phonon losses are discussed and estimated.

\end{abstract}

\maketitle

\section*{Introduction}

A Fabry-P\'{e}rot interferometer is an optical resonant system that typically consists of two parallel highly reflecting mirrors.
Due to their high quality factors, a Fabry-P\'{e}rot cavity is often used to store photons for almost milliseconds while they bounce between the mirrors. Low loss is necessary in order to increase interaction between light and matter or other physical substances under study. Such systems are widely used in telecommunication, lasers, gravitational wave detection, spectroscopy, etc.

Similar principles can be applied to the field of acoustics, bulk phonons may be resonantly confined in a thin film or a crystal lattice to build a phonon cavity. In this case, resonant phonons are stored between borders that separate different media playing a role of mirrors.  

Losses in acoustic resonators can be classified into two groups: first, losses attributed to intrinsic material processes, second, losses due to resonator design. The latter are mostly due to energy leakage through clamping, substrates, electrodes, etc~\cite{Aspelmeyer}. The goal of the cavity design is to make "engineering losses" negligible in order to achieve a regime where losses are due to the material itself. If this regime is achieved the cavity quality factor can be increased by orders of magnitude by cooling to cryogenic temperatures.

 The thin film approach (realised in Thin Film Bulk Acoustic Resonators and High Overtone Bulk Acoustic Resonators) provides technology that can guarantee smaller distances between the borders, and thus higher resonant frequencies. The main disadvantages of these systems are the need for a substrate and the inability to efficiently control the film geometry due to limitations of the deposition process. Both result in additional losses due to phonon leakage into the environment.   
 
The crystalline lattice approach utilises structures built from pieces of single crystal material (e.g. quartz Bulk Acoustic Wave (BAW) resonator). Due to a completely different fabrication process, this technology can provide a more sophisticated design that guarantees sufficient limitation of phonon leakage. In addition, extremely low losses of such devices extend the possible frequency range to comparatively high frequencies. In this work, we show that this approach allows $Q$-factors of above 1 billion with a maximum of 8 billion corresponding to a $Q\times f$ product of $1.6\cdot10^{18}$ at liquid helium temperatures. This is 310 times higher than the previous value achieved at this temperature~\cite{galliou:091911}. Such high sensitivity should allow future applications of acoustical cavities to be broadened and be used as an alternative (to optical cavities) experimental base for broad range of physical experiments, including material characterisation~\cite{Luthi}, quantum cavity electromechanics~\cite{Mahboob, Ruskov}, quantum metrology~\cite{Kippen,aspelmayer2} etc.  


\section*{Results}
\subsection*{Phonon Excitation and Trapping}

Vibrations of ideal plate resonators have been studied for many decades~\cite{Mindlin2007}. Unlike their idealised counterparts, actual cavities pose two almost contrary requirements. First, they have to be coupled to the environment for effective excitation and measurements, second, they must be uncoupled from the environment in order to avoid phonon leakage. The solution to this corundrum is to excite vibrations with the electro-magnetic field while mechanically separating the crystal from the environment.

 Coupling to electromagnetic measuring and excitation fields is introduced by making a vibrating body from piezoelectric material. Imposing a potential difference between two surfaces of the plate creates {a gradient of {the electric potential} along the plate thickness. The spatial Fourier harmonics of this gradient are coupled to mechanical vibrations {through the} piezoelectric effect. Frequencies of these vibrations are set by plate dimensions, material properties and the order of the harmonic. This type of vibration is referred as a thickness vibration that has to be distinguish from parasitic flexure vibration modes. Detection of the plate vibration is performed by measuring the displacement current between the surfaces of the plate. }
 
Due to the surface boundary conditions, such a cavity can sustain resonant phonons of discrete wavelengths. When describing an acoustic mode in a piezoelectric plate, the number of half-waves in the plate thickness is referred as the overtone (OT) number, and the resonance itself as an overtone. The cavity geometry allows only anti-symmetric types of vibration {to be excited piezoelectrically, thus limiting experimentally observed} OTs only to odd numbers. 

In order to avoid any energy leakage, the vibrating part of the plate has to be mechanically separated from the environment. This is achieved by two means: first, the excitation electrodes are separated from the crystal plate by a tiny vacuum gap; second, acoustic energy is trapped in the centre of the disk plate while the disk is clamped from the sides. The trapping is done by introducing curvature into one of the plate surfaces by manufacturing a plano-convex shape~\cite{stevens:1811,Goryachev1}. This variable plate thickness increases the local phonon frequency along the plate disk radius closer to the edges. The corresponding energy difference along the radial direction due to this curvature can be considered as an effective potential well. So, phonons are trapped if their total transverse energy (corresponding to in-plane wave numbers) do not exceed the energy difference between the edge and the centre. This situation significantly decreases the probability to find a resonant phonon near the clamping points. Theoretical treatment of the energy trapping problem from the point of view of linear elastodynamics is given in~\cite{stevens:1811}. { The calculations predict the mode shape in the form of polynomial-normal distribution. This result has been confirmed by means of X-ray spectroscopy~\cite{Tanner, slavov}.}


\begin{figure}[t!]
\centering
\includegraphics[width=3.25in]{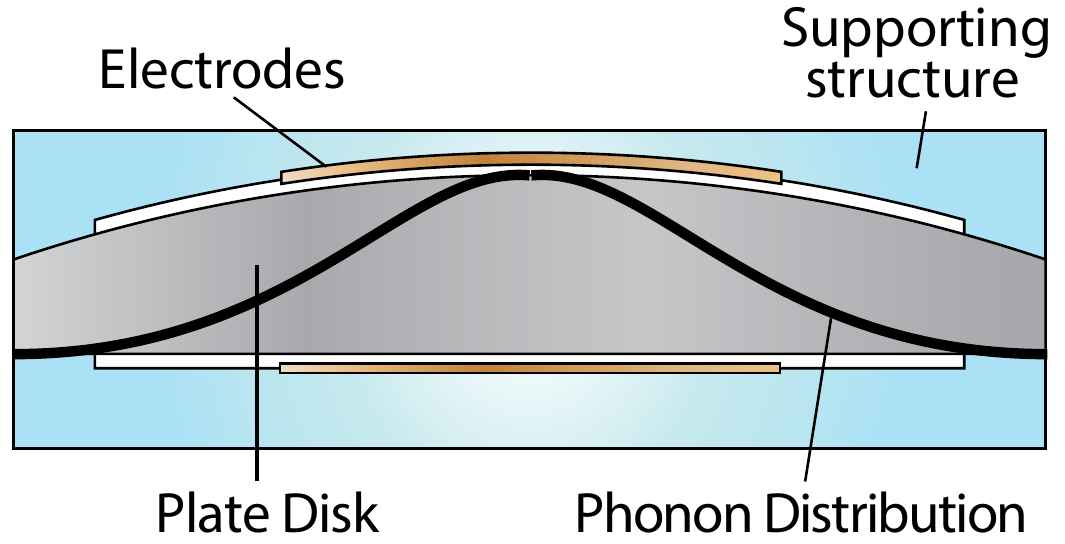}
\caption{\label{resonator}Cross section in the $r-z$ plane of the quartz Bulk Acoustic Wave (BAW) resonator of cylindrical geometry with typical distribution of trapped phonons in the radial direction.}
\end{figure}

\subsection*{Details of Cavity Design}

Tested acoustic cavities are made from Top High Quality (THQ) alpha synthetic quartz produced by Gemma Quartz \& Crystal according to a hydrothermal growth process. Impurities are removed from the heart of the quartz bars before the latter are cut up in oriented blanks.  Disks come from slices cut out from a synthetic quartz crystal bar according to specific angles with respect to the crystalline axis (so-called singly rotated cuts or doubly rotated cuts). The tested cavities are doubly-rotated Stress Compensated (SC) cut plates~\cite{1536996}, which is one of the most commonly used cuts. Doubly rotated cut cavities support three possible phonon polarizations with different velocities: the quasi-longitudinal ($V_L = 6757$ m/s), fast quasi-shear ($V_{FS} = 3966$ m/s) and slow quasi-shear ($V_{SS} = 3611$ m/s). All three types of polarisation are not pure due to crystal anisotropy. According to the contoured-resonator axis $[x, y, z]$~\cite{IEEE176}, the y-axis being normal to the plate, their polarizations are respectively: $[0.226, 0.968, 0.111]$, $[-0.198, -0.0657, 0.979]$, $[0.957, -0.243, 0.177]$.

After the slices are cut they are then rounded, lapped and polished resulting in plano-convex contoured plates. Polishing is a key step of the process: it consists of a mixing of mechanical polishing and chemical etching. This last operation is mandatory to remove the "dead layer", i.e. a damaged layer - a blend of amorphous and polycrystalline randomly-oriented matter resulting from the first process. Chemical etching is also one efficient way to control precisely the thickness of the resonator, a key parameter for defining the frequency of waves propagating along the thickness axis. At least, mechanical strains induced by machining should also be removed by an additional reheat.

The cavity design for which both electrodes are not deposited on the vibrating plate itself is known as (BVA)~\cite{1537081}. Electrodes are deposited on two quartz supports placed on each side of the resonator, and in such a way that the gap between electrodes and resonator surfaces is typically between $5$ and $10 \mu$m~\cite{galliou:091911,Goryachev1}. 

In this work we compare performance of the following types of acoustical cavities:
\begin{enumerate}
\item $1.08$ mm thick, $13$ mm diameter electrode-separated disk cavities initially designed to sustain shear vibration of $5$~MHz at room temperature (manufactured by BVA Industrie)~{\cite{galliou:091911}};
\item $0.54$ mm thick, $13$ mm diameter electrode-separated disk cavities initially designed to sustain shear vibration of $10$~MHz at room temperature (manufactured by BVA Industrie)~{~\cite{Goryachev1}};
\item $1$ mm thick, $30$ mm diameter electrode-separated disk cavities with higher grade surface polishing initially designed to sustain shear vibration of $5$~MHz at room temperature  (manufactured by Oscilloquartz SA);
\item $0.54$ mm thick, $13$ mm diameter electrode-deposited disk cavities initially designed to sustain shear vibration of $10$~MHz at room temperature  (manufactured at  FEMTO-ST).
\end{enumerate}

It has to be noted {that all cavities are initially designed for room temperature operation with slow-speed thickness-shear vibrations and not for longitudinally-polarized ones. This is due to the fact that the frequency-temperature characteristic of the former exhibits a turnover point, which is not the case of the latter. Moreover} at liquid helium temperatures, longitudinal phonons exhibit significantly lower losses than any shear phonons~\cite{galliou:091911}. Direct cavity optimisation for liquid helium operation is still impossible due to unknown temperature coefficients of the elastic constants at these temperatures.

\begin{figure}[t!]
\centering
\includegraphics[width=3.25in]{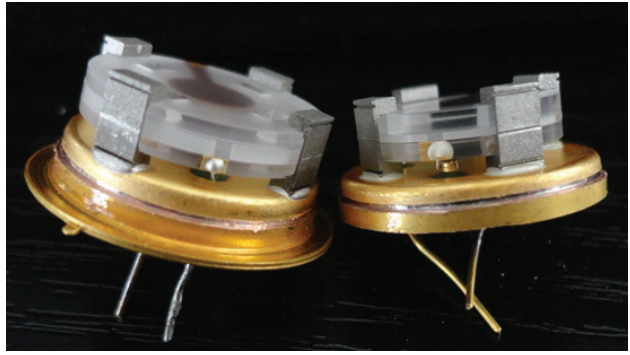}
\caption{\label{resonator}Photograph of the electrode and cavity mounting arrangement of the quartz BAW cavities.}
\end{figure}

It has to be noted that two parallel excitation electrodes form a parasitic capacitor (typically close to $3$~pF)  creating a parallel photon branch. At higher frequencies, the number of these non-resonant photons exceeds the number of acoustic phonons decreasing the mode contrast.  Hence, the parasitic capacitance uncouples the acoustic motion from the electromagnetic environment. Nevertheless, the excitation of phonons of shorter wavelengths is possible with effective electrical compensation of this shunt capacitance.


\section*{Discussion}

\begin{figure}[t!]
\centering
\includegraphics[width=3.25in]{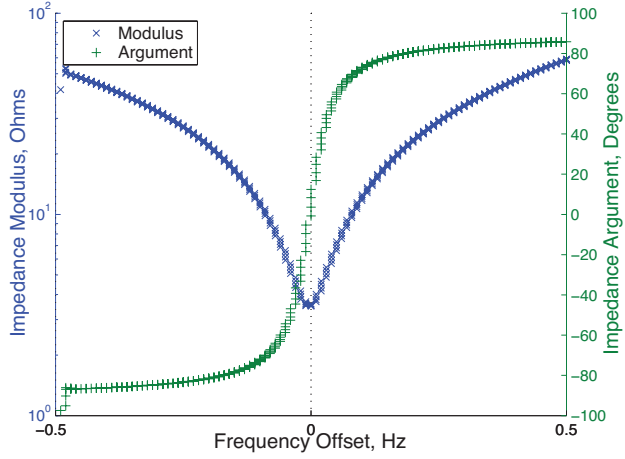}
\caption{\label{admit} Measured impedance (absolute blue and phase) of the 5th OT of the longitudinal mode ($15.7$~MHz) at $3.75$ K for the cavity number 3.}
\end{figure}

The four cavities described above were characterised for resonant longitudinal phonons of {half} wavelengths varying from $\frac{1}{5}$ to $\frac{1}{77}$ of the plate thickness. { $Q$-factors were extracted by fitting measured electric admittance of the resonator around each resonance frequency. Fig.~\ref{admit} shows an example of experimental results for the 3rd type of acoustic cavities measured for the 5th OT at $3.75$K.} 

\begin{figure}[t!]
\centering
\includegraphics[width=3.45in]{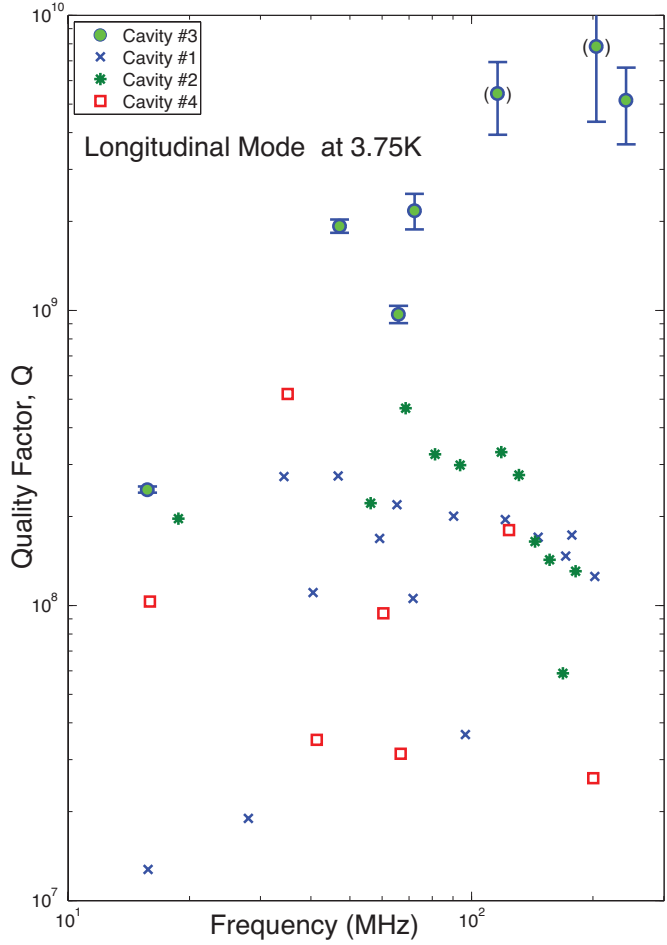}
\caption{\label{QS_New_nogrid} Measured quality factors for different OTs of the (quasi-)longitudinal vibration of the four cavities under investigation at $3.75$ K. Error bars for standard deviations lower than $\pm2.5$\% are not represented. Both higher values in brackets are close to the measurement set-up limitation.}
\end{figure}

\begin{figure}[t!]
\centering
\includegraphics[width=3.45in]{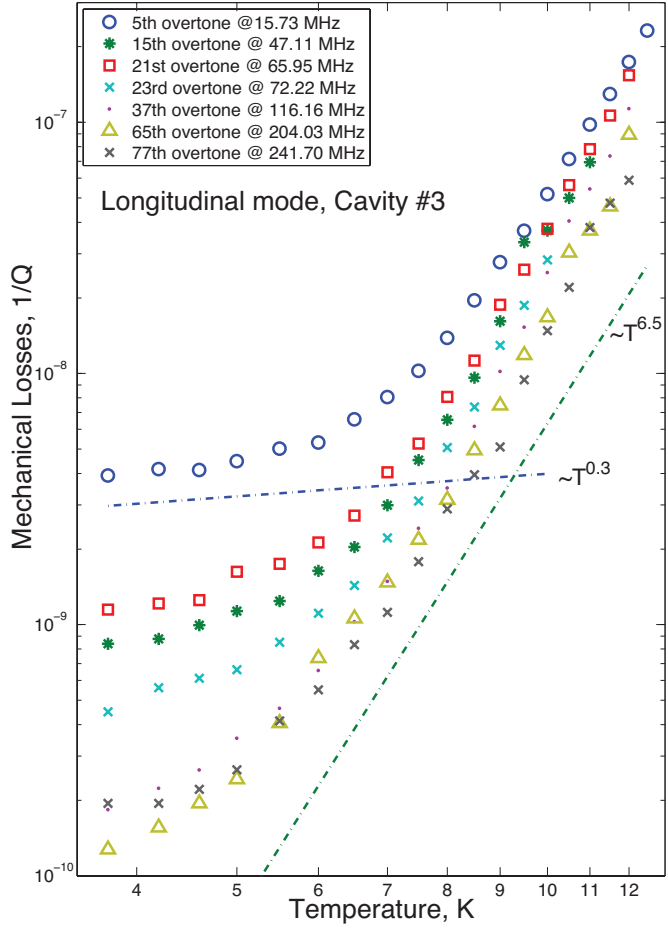}
\caption{\label{TS_New_nogrid} Temperature dependance of cavity losses for different OTs of the (quasi-)longitudinal vibration in the cavity number 3. }
\end{figure}

The results (see Fig.~\ref{QS_New_nogrid}) show that the cavity with the largest diameter and better polishing exhibits significantly lower losses. This fact could be attributed to better trapping (due to larger potential well) and lower scattering on surface imperfections. In addition, all four samples exhibited an increase of $Q$ as a function of OT number, at least for a certain range of frequencies. This is explained by improving of the phonon trapping with increasing OT number as follows from the theory~\cite{stevens:1811}.  
Temperature dependence of the selected overtones for this cavity is shown in Fig.~\ref{TS_New_nogrid}. 

Total intrinsic losses of an acoustic cavity apart from lack of trapping are a sum of losses due to different mechanisms:
\begin{equation}
\label{loss}
\left. \begin{array}{ll}
\displaystyle\frac{1}{Q} = \frac{1}{Q_{phonon-phonon}}+\frac{1}{Q_{TLS}}+\\
\displaystyle\hspace{35pt}+\frac{1}{Q_{scattering}}+\frac{1}{Q_{thermoelastic}}+\text{etc},
\end{array} \right. 
\end{equation}
where the $Q$-values represent different mechanisms.  { These mechanisms are explained below combining the results of frequency (Fig.~\ref{QS_New_nogrid}) and temperature (Fig. \ref{TS_New_nogrid}) dependences}. 

{ First,} the most important mechanism of energy loss in acoustic cavities at low temperatures and relatively high frequencies is loss mechanism due to phonon-phonon scattering over crystal lattice anharmonicity {($Q_{phonon-phonon}$ in equation~(\ref{loss}))}. This mechanism {was} first analytically considered by Landau for the case of shear-polarized phonons~\cite{landaurumer1,landaurumer2}. It was also pointed out that longitudinal phonons do not exhibit this type of losses in the case of two phonon collision, due to significantly higher sound velocity. Later the theory was extended to the case of three colliding phonons~\cite{Maris1971} predicting  the following law: $Q_{phonon-phonon}^{-1}\sim T^{-6.5}$. This law is experimentally observed for the large acoustic cavity between between $6$K and $13$K { (Fig. \ref{TS_New_nogrid})}.
According to the theory~\cite{Maris1971}, the acoustic wave attenuation $\alpha$ is proportional to the resonance angular frequency $\omega$ as long as $\frac{\hbar \omega}{k_BT} << 1$, and $\omega \tau \gg 1$ where $\tau$ denotes the average lifetime of thermal phonons, $\hbar$ the Planck constant, and $k_B $ the Boltzmann constant. As a consequence,
the $Q_{phonon-phonon}$-factor of an acoustic wave propagating with velocity $V$
\begin{equation}
\label{basic_Q}
Q = \frac{\pi\omega}{V\alpha}
\end{equation}
would no longer depend on frequency $\omega$. Since these conditions are met in the present experiment, the Q-factor tends to be independent of $f$ ($Q=const$ law) as shown in Fig.~\ref{QS_New_nogrid}.

 { Second,} for temperatures below $6$K the $Q(T)$ dependence changes the slope tending to $T^{-0.3}$ scaling law { (Fig. \ref{TS_New_nogrid})}. Such dependence have been observed in a various of electromechanical systems \cite{PhysRevLett.93.185501,doi:10.1021/nl900612h,shim:133505,PhysRevB.72.224101,PhysRevB.79.125424}, { and acoustic cavities at lower temperatures~\cite{Goryachev1}, in particular.} The $T^{-1/3}$ dependence is theoretically explained by coupling to  so-called ``two level systems" (TLS)~\cite{0295-5075-78-6-60002} {($Q_{TLS}$ in equation~(\ref{loss})}). { In synthetic quartz these TLSs are attributed to impurity ions, typically $Al^{3+}$, $N^+$, $Li^+$, $Si^{4+}$ and $H^+$, etc~\cite{Martin, MasonWP, Euler,Halliburton} studied by electron-spin resonance, infrared spectroscopy, dielectric relaxation spectroscopy and thermoluminescence. Although the utilised material is premium pure, concentration of these ions could reach $1-3$~ppb.} 

 { Third,} the increase of cavity losses at higher frequencies occurs for the most of resonators (see Fig.~\ref{QS_New_nogrid}), and can be explained by phonon scattering on the cavity surfaces ($Q_{scattering}$ in equation~(\ref{loss})). Scattering originates from the surface roughness as well as from the presence of a Òdead layerÓ. Indeed, the reflection coefficient on a rough surface with a roughness variance $\sigma^2$, assuming a gaussian probability density, and can be expressed as~\cite{Welton1973} $R=e^{-2k^2\sigma^2}$, where $k$ is the wave number.
At the same time, propagating through the thickness $h$, the acoustic wave is attenuated by $e^{-\alpha h}$.
Thus, the reflection is equivalent to a modified path causing an attenuation equal to $\alpha^{'} = \alpha + ln(R/h)$. Consequently, for an OT number $n$ such that $kh=n\pi$, the associated $Q$-factor becomes:
\begin{equation}
\label{routh}
Q_{scattering} = \frac{h^2}{2n\sigma^2}
\end{equation}
Since the resonant frequency $\omega$ is almost proportional to the overtone number $n$, this mechanism results in $Q\times \omega = const$ law. 
{ The surface scattering mechanism strongly depends on the polishing quality denoted by the roughness variance $\sigma$. This explains the better performance of cavity number {3}  (Fig. \ref{TS_New_nogrid}), which has a better surface polishing grading. The largest observed values of quality factor were at $n=65$ for cavity number {3}, which has a $\sigma$ of one $nm$ (consistent with specification of Oscilloquartz fabrication process). For this type of cavity, $Q_{scattering}$ is estimated to be $\frac{2}{n}\cdot10^{12}$, which gives an upper limit of $3\times10^{10}$ for the highest measured OT $n = 65$ at $204$~MHz.
For other cavities the polishing process is inferior with $\sigma$ varying from $4$ to $7$~nm. For example, for cavity type 2 with a specified $\sigma = 4$~nm, $Q_{scattering}$ is found to be $\frac{8}{n}\cdot10^9$ and gives a value of $4.2\times10^8$ at $n=19$ ($118.6$ MHz), which is consistent with the experimentally measured value of $3.3\times 10^8$.
Thus, we can conclude that the highest measured $Q$-values are close to estimated values of $Q_{scattering}$,  demonstrating that the cavity roughness could be a major factor limiting the Q-factor for high frequencies. }

 { Fourth,} although thermoelastic losses do not exist for pure shear vibration, since there is no volume change, they are present for the longitudinal modes. {So, the $Q_{thermoelastic}$ for longitudinal vibration can be estimated based on the description of the wave propagation including the coupled equations of motion of the medium and thermal conduction~\cite{Deresiewicz1954}:
\begin{eqnarray}
\label{motion_thermic}
\rho\ddot{u_{i}} = c_{ijkl}u_{k,lj} - \lambda_{ij}\delta T_{,j} \\
\Gamma_{ij}T_{,ij}- \rho C^{S} \dot{T} - T_{0} \lambda_{kl} u_{k,l}=0  
\end{eqnarray}
with $\lambda_{ij} = c_{ijkl}\alpha_{kl}$, and where $u_{i}$ is the displacement vector, $c_{ijkl}$ the tensor of isothermal elastic constants, $\alpha_{ij}$ the thermal expansion tensor, $\Gamma_{ij}$ the thermal conductivity tensor, $\rho$ the volumic mass, and $C^{S}$ the specific heat at constant strain, $\delta T$ the temperature change around the reference temperature $T_{0}$. Although, most of these coefficients for the lightly anisotropic quartz are unknown at cryogenic conditions, 
simplified calculations can be made assuming an isotropic medium and propagation along the $Z$-axis. The optical $Z$-axis is the most experimentally studied case.}
This analysis implies calculation of a turning-point frequency $f^\ast$ that shows the regime of acoustic wave propagation - adiabatic or {isothermal:
\begin{equation}
\label{fast}
f^\ast = \frac{V_{3}^2}{2\pi\chi_{3}}
\end{equation}
where $\chi_{3} = \frac{\Gamma_{33}}{\rho C^{S}}$ is the material diffusivity, and $ V_{3}= \sqrt{\frac{c_{33}}{\rho}}$ for a propagation along the $Z$-axis}. The former is calculated based on data on thermal conductivity and heat capacity at cryogenic temperatures~\cite{Barron1982,Zeller1971} giving $1 \leq f^\ast \leq 5.5$~MHz for $3 \leq T \leq 10$K. Over this range, all the resonant frequencies are always greater than $f^\ast$, and acoustic wave propagates in the quasi-isothermal regime. {For this regime, the maximum attenuation is:
\begin{equation}
\label{max_attenuation}
\alpha_{\infty}=\frac{\epsilon_{3} \pi f^\ast}{V_{3}}
\end{equation}
with $\epsilon_{3}=\frac{\lambda_{33}^{2}T}{\rho C^{S}c_{33}}$, where $\lambda_{33}=2c_{12}\alpha_{11}+c_{33}\alpha_{33}$, because $\alpha_{11} = \alpha_{22}$ and $c_{13} = c_{23}$ in case of crystalline quartz. Whereas the thermal expansion coefficients have been measured at low temperatures~\cite{Barron1982},the elastic constants on the other hand can just be roughly approximated by their values at room temperature. Nevertheless, the corresponding thermoelastic losses estimated from equation~(\ref{basic_Q}) remain always at least two orders of magnitude lower than those measured below $10$K.} Since $\alpha_{\infty}$ is almost independent of frequency for a given material, these type of losses result in $Q/\omega=const$ law. 
 


We demonstrate that acoustic cavities with effective trapping of longitudinally polarised phonons can exhibit quality factors above 1 Billion at liquid helium temperatures and frequencies $10-200$ MHz. This shows that phonon cavities can be considered as an alternative to photon cavities in many low-frequency, low-power experiments. The advantages of acoustic devices is the ease of operation and integration into higher-order electronic systems, compact size, natural way of combination of mechanical motion and electromagnetic environment, as well as artificial lattice impurities. 
 In particular, these systems can be used as parts of quantum hybrid systems in miliKelvin temperatures integrating quantum circuit electrodynamics and cavity elastodynamics. Moreover, unlike nano-scale mechanical resonators, acoustic cavities are a macroscopic objects of considerable {effective} mass varying 
 { in the range of $0.1-10$ milligrams for modes between $100$ and $10$~MHz.}
This, can be a promising experiment on unifying quantum mechanics and general relativity in a single experiment~\cite{aspelmayer2}. 

Apart from fundamental physics, cryogenic acoustic cavities can be used for building ultra-stable frequency sources for variety of applications.  

\section*{Methods}

The device under test is embedded inside an oxygen-free copper block fixed on the end of the cold finger of a Helium pulse-tube cryocooler~\cite{galliou:091911}. The cold finger is temperature-controlled, and residual fluctuations of temperature are within $\pm2$mK around the setting temperature.

The device characterisation is performed using the impedance analyzis method: measurements of the displacement current through the resonator thickness while applying alternating potential difference between electrodes. Their ratio is the complex impedance that is measured for a range of frequencies around a resonance. This characteristic is used to calculate a phonon resonance bandwidth. For accurate measurements of intrinsic properties of a device under test (DUT), the method requires a calibration procedure in order to compensate any influence of a long connecting cables. 
For this purpose, three identical extra cables follow the same path to be ended respectively by an open-circuit, a short circuit and a $50$ Ohms load.
Measurements of these etalons helps to cancel cable influence for certain values of the analyzer output power, frequency span and sweep time.
 
 The excitation power is kept as low as possible in order to avoid nonlinear and thermal effects. 
Power dissipation inside the resonator is estimated to vary from $0.1$ nW to $3$ nW, depending on a mode coupling and quality factor. { Nonlinear effects of the investigated devices are due to lattice anharmonicity. They strongly depend on quality factor of a mode and can be noticeable at powers as low as $10$~nW. }

Special attention should be paid to the accuracy when characterising modes with extremely high quality factors due to extreme narrowness of bandwidths (below $100$~mHz). The frequency span of the analyzer is adjusted to get the maximum benefit of the number $N$ of measurement points. The minimum available span of $0.5$ Hz for 401 measured points provides resolution of about $1.25$~{mHz}. The minimum sweep time required to avoid all transient effects is $(N-1)\times\tau$ where $\tau$ is phonon relaxation time of a mode of {frequency $f$}. It is given as $\tau = \frac{Q_L}{\pi f}$ where $Q_L$ is a mode loaded quality factor, $f$ is the resonant frequency. 

Based on the  specifications of the measurement equipment and  the post-processing procedure, the measurement uncertainty {of the intrinsic (unloaded) quality-factor $Q$} is evaluated as:
\begin{equation}
\label{uncert}
\frac{dQ}{Q}\approx 2Q\cdot \frac{df}{f},
\end{equation}
which demonstrates importance of the fractional frequency stability of both the measurement apparatus frequency source and the DUT itself. The latter stability is improved by locking to an ultra-stable frequency source, such as a Hydrogen maser. The long term (due to requirement for large sweep time) frequency stability of the resonator mode is determined predominantly by its temperature sensitivity and $\frac{df}{f}$ does not typically exceed $10^{-10}$ providing measurement uncertainty below a few percent for all characterised modes.

\section*{Author Contributions}
SG and MG made cavity characterisation. PA participated in organising cryogenic experiments. JPA prepared cavity 4 and consulted on its design. RB consulted on cavity design and manufacturing processes. SG, MG, RB and MT discussed loss mechanisms. MG proposed the terminology.  
SG, MG and MT wrote the main manuscript text. MT supervised the work. All authors reviewed the manuscript.

\section*{Competing Financial Interests}
This work was supported by the Australian Research Council Grant No. CE11E0082 and FL0992016, {and the Regional Council of Franche Comt\'{e}, France, Grant No. 2008C16215}.

\section*{References}

\begin{thebibliography}{32}%
\makeatletter
\providecommand \@ifxundefined [1]{%
 \@ifx{#1\undefined}
}%
\providecommand \@ifnum [1]{%
 \ifnum #1\expandafter \@firstoftwo
 \else \expandafter \@secondoftwo
 \fi
}%
\providecommand \@ifx [1]{%
 \ifx #1\expandafter \@firstoftwo
 \else \expandafter \@secondoftwo
 \fi
}%
\providecommand \natexlab [1]{#1}%
\providecommand \enquote  [1]{``#1''}%
\providecommand \bibnamefont  [1]{#1}%
\providecommand \bibfnamefont [1]{#1}%
\providecommand \citenamefont [1]{#1}%
\providecommand \href@noop [0]{\@secondoftwo}%
\providecommand \href [0]{\begingroup \@sanitize@url \@href}%
\providecommand \@href[1]{\@@startlink{#1}\@@href}%
\providecommand \@@href[1]{\endgroup#1\@@endlink}%
\providecommand \@sanitize@url [0]{\catcode `\\12\catcode `\$12\catcode
  `\&12\catcode `\#12\catcode `\^12\catcode `\_12\catcode `\%12\relax}%
\providecommand \@@startlink[1]{}%
\providecommand \@@endlink[0]{}%
\providecommand \url  [0]{\begingroup\@sanitize@url \@url }%
\providecommand \@url [1]{\endgroup\@href {#1}{\urlprefix }}%
\providecommand \urlprefix  [0]{URL }%
\providecommand \Eprint [0]{\href }%
\@ifxundefined \urlstyle {%
  \providecommand \doi  [0]{\begingroup \@sanitize@url \@doi}%
  \providecommand \@doi [1]{\endgroup \@@startlink {\doibase
  #1}doi:\discretionary {}{}{}#1\@@endlink }%
}{%
  \providecommand \doi  [0]{doi:\discretionary{}{}{}\begingroup
  \urlstyle{rm}\Url }%
}%
\providecommand \doibase [0]{http://dx.doi.org/}%
\providecommand \Doi [0]{\begingroup \@sanitize@url \@Doi }%
\providecommand \@Doi  [1]{\endgroup\@@startlink{\doibase#1}\@@Doi}%
\providecommand \@@Doi [1]{#1\@@endlink}%
\providecommand \selectlanguage [0]{\@gobble}%
\providecommand \bibinfo  [0]{\@secondoftwo}%
\providecommand \bibfield  [0]{\@secondoftwo}%
\providecommand \translation [1]{[#1]}%
\providecommand \BibitemOpen [0]{}%
\providecommand \bibitemStop [0]{}%
\providecommand \bibitemNoStop [0]{.\EOS\space}%
\providecommand \EOS [0]{\spacefactor3000\relax}%
\providecommand \BibitemShut  [1]{\csname bibitem#1\endcsname}%
\bibitem [{\citenamefont {Cole}\ \emph {et~al.}(2011)\citenamefont {Cole},
  \citenamefont {Wilson-Rae}, \citenamefont {Werbach}, \citenamefont {Vanner},\
  and\ \citenamefont {Aspelmeyer}}]{Aspelmeyer}%
  \BibitemOpen
  \bibfield  {author} {\bibinfo {author} {\bibfnamefont {G.}~\bibnamefont
  {Cole}}, \bibinfo {author} {\bibfnamefont {I.}~\bibnamefont {Wilson-Rae}},
  \bibinfo {author} {\bibfnamefont {K.}~\bibnamefont {Werbach}}, \bibinfo
  {author} {\bibfnamefont {M.}~\bibnamefont {Vanner}}, \ and\ \bibinfo {author}
  {\bibfnamefont {M.}~\bibnamefont {Aspelmeyer}},\ }\bibfield  {title}
  {\enquote {\bibinfo {title} {Phonon-tunnelling dissipation in mechanical
  resonators},}\ }\href@noop {} {\bibfield  {journal} {\bibinfo  {journal}
  {Nature Communications},\ }\textbf {\bibinfo {volume} {2}} (\bibinfo {year}
  {2011})}\BibitemShut {NoStop}%
\bibitem [{\citenamefont {Galliou}\ \emph {et~al.}(2011)\citenamefont
  {Galliou}, \citenamefont {Imbaud}, \citenamefont {Goryachev}, \citenamefont
  {Bourquin},\ and\ \citenamefont {Abbe}}]{galliou:091911}%
  \BibitemOpen
  \bibfield  {author} {\bibinfo {author} {\bibfnamefont {S.}~\bibnamefont
  {Galliou}}, \bibinfo {author} {\bibfnamefont {J.}~\bibnamefont {Imbaud}},
  \bibinfo {author} {\bibfnamefont {M.}~\bibnamefont {Goryachev}}, \bibinfo
  {author} {\bibfnamefont {R.}~\bibnamefont {Bourquin}}, \ and\ \bibinfo
  {author} {\bibfnamefont {P.}~\bibnamefont {Abbe}},\ }\bibfield  {title}
  {\enquote {\bibinfo {title} {Losses in high quality quartz crystal resonators
  at cryogenic temperatures},}\ }\href@noop {} {\bibfield  {journal} {\bibinfo
  {journal} {Applied Physics Letters},\ }\textbf {\bibinfo {volume} {98}},\
  \bibinfo {pages} {091911} (\bibinfo {year} {2011})}\BibitemShut {NoStop}%
\bibitem [{\citenamefont {Luthi}(2005)}]{Luthi}%
  \BibitemOpen
  \bibfield  {author} {\bibinfo {author} {\bibfnamefont {B.}~\bibnamefont
  {Luthi}},\ }\href@noop {} {\emph {\bibinfo {title} {Physical Acoustics in
  Solid State Science}}},\ Solid-State Sciences\ (\bibinfo  {publisher}
  {Springer-Verlag},\ \bibinfo {address} {Berlin},\ \bibinfo {year}
  {2005})\BibitemShut {NoStop}%
\bibitem [{\citenamefont {Mahboob}\ \emph {et~al.}(2012)\citenamefont
  {Mahboob}, \citenamefont {Nishiguchi}, \citenamefont {Okamoto},\ and\
  \citenamefont {Yamaguchi}}]{Mahboob}%
  \BibitemOpen
  \bibfield  {author} {\bibinfo {author} {\bibfnamefont {I.}~\bibnamefont
  {Mahboob}}, \bibinfo {author} {\bibfnamefont {K.}~\bibnamefont {Nishiguchi}},
  \bibinfo {author} {\bibfnamefont {H.}~\bibnamefont {Okamoto}}, \ and\
  \bibinfo {author} {\bibfnamefont {H.}~\bibnamefont {Yamaguchi}},\ }\bibfield
  {title} {\enquote {\bibinfo {title} {Phonon-cavity electromechanics},}\
  }\href@noop {} {\bibfield  {journal} {\bibinfo  {journal} {Nature Physics},\
  }\textbf {\bibinfo {volume} {8}},\ \bibinfo {pages} {387} (\bibinfo {year}
  {2012})}\BibitemShut {NoStop}%
\bibitem [{\citenamefont {Ruskov}\ and\ \citenamefont {Tahan}(2012)}]{Ruskov}%
  \BibitemOpen
  \bibfield  {author} {\bibinfo {author} {\bibfnamefont {R.}~\bibnamefont
  {Ruskov}}\ and\ \bibinfo {author} {\bibfnamefont {C.}~\bibnamefont {Tahan}},\
  }\bibfield  {title} {\enquote {\bibinfo {title} {On-chip quantum
  phonodynamics},}\ }\href@noop {} {\bibfield  {journal} {\bibinfo  {journal}
  {http://arxiv.org/abs/1208.1776}} (\bibinfo {year} {2012})}\BibitemShut
  {NoStop}%
\bibitem [{\citenamefont {Aspelmeyer}\ \emph {et~al.}(2013)\citenamefont
  {Aspelmeyer}, \citenamefont {Kippenberg},\ and\ \citenamefont
  {Marquardt}}]{Kippen}%
  \BibitemOpen
  \bibfield  {author} {\bibinfo {author} {\bibfnamefont {M.}~\bibnamefont
  {Aspelmeyer}}, \bibinfo {author} {\bibfnamefont {T.}~\bibnamefont
  {Kippenberg}}, \ and\ \bibinfo {author} {\bibfnamefont {F.}~\bibnamefont
  {Marquardt}},\ }\bibfield  {title} {\enquote {\bibinfo {title} {Cavity
  optomechanics},}\ }\href@noop {} {\bibfield  {journal} {\bibinfo  {journal}
  {http://arxiv.org/abs/1303.0733}} (\bibinfo {year} {2013})}\BibitemShut
  {NoStop}%
\bibitem [{\citenamefont {Pikovski}\ \emph {et~al.}(2012)\citenamefont
  {Pikovski}, \citenamefont {Vanner}, \citenamefont {Aspelmeyer}, \citenamefont
  {Kim},\ and\ \citenamefont {Brukner}}]{aspelmayer2}%
  \BibitemOpen
  \bibfield  {author} {\bibinfo {author} {\bibfnamefont {I.}~\bibnamefont
  {Pikovski}}, \bibinfo {author} {\bibfnamefont {M.}~\bibnamefont {Vanner}},
  \bibinfo {author} {\bibfnamefont {M.}~\bibnamefont {Aspelmeyer}}, \bibinfo
  {author} {\bibfnamefont {M.}~\bibnamefont {Kim}}, \ and\ \bibinfo {author}
  {\bibfnamefont {C.}~\bibnamefont {Brukner}},\ }\bibfield  {title} {\enquote
  {\bibinfo {title} {Probing planck-scale physics with quantum optics},}\
  }\href@noop {} {\bibfield  {journal} {\bibinfo  {journal} {Nature Physics},\
  }\textbf {\bibinfo {volume} {8}},\ \bibinfo {pages} {393} (\bibinfo {year}
  {2012})}\BibitemShut {NoStop}%
\bibitem [{\citenamefont {Mindlin}(2007)}]{Mindlin2007}%
  \BibitemOpen
  \bibfield  {author} {\bibinfo {author} {\bibfnamefont {R.}~\bibnamefont
  {Mindlin}},\ }\href@noop {} {\emph {\bibinfo {title} {An Introduction to the
  Mathematical Theory of Vibrations of Elastic Plates}}},\ edited by\ \bibinfo
  {editor} {\bibfnamefont {J.}~\bibnamefont {Yang}}\ (\bibinfo  {publisher}
  {World Scientific},\ \bibinfo {year} {2007})\BibitemShut {NoStop}%
\bibitem [{\citenamefont {Stevens}\ and\ \citenamefont
  {Tiersten}(1986)}]{stevens:1811}%
  \BibitemOpen
  \bibfield  {author} {\bibinfo {author} {\bibfnamefont {D.~S.}\ \bibnamefont
  {Stevens}}\ and\ \bibinfo {author} {\bibfnamefont {H.~F.}\ \bibnamefont
  {Tiersten}},\ }\bibfield  {title} {\enquote {\bibinfo {title} {An analysis of
  doubly rotated quartz resonators utilizing essentially thickness modes with
  transverse variation},}\ }\href@noop {} {\bibfield  {journal} {\bibinfo
  {journal} {The Journal of the Acoustical Society of America},\ }\textbf
  {\bibinfo {volume} {79}},\ \bibinfo {pages} {1811} (\bibinfo {year}
  {1986})}\BibitemShut {NoStop}%
\bibitem [{\citenamefont {Goryachev}\ \emph {et~al.}(2012)\citenamefont
  {Goryachev}, \citenamefont {Creedon}, \citenamefont {Ivanov}, \citenamefont
  {Galliou}, \citenamefont {Bourquin},\ and\ \citenamefont
  {Tobar}}]{Goryachev1}%
  \BibitemOpen
  \bibfield  {author} {\bibinfo {author} {\bibfnamefont {M.}~\bibnamefont
  {Goryachev}}, \bibinfo {author} {\bibfnamefont {D.~L.}\ \bibnamefont
  {Creedon}}, \bibinfo {author} {\bibfnamefont {E.~N.}\ \bibnamefont {Ivanov}},
  \bibinfo {author} {\bibfnamefont {S.}~\bibnamefont {Galliou}}, \bibinfo
  {author} {\bibfnamefont {R.}~\bibnamefont {Bourquin}}, \ and\ \bibinfo
  {author} {\bibfnamefont {M.~E.}\ \bibnamefont {Tobar}},\ }\bibfield  {title}
  {\enquote {\bibinfo {title} {Extremely low-loss acoustic phonons in a quartz
  bulk acoustic wave resonator at millikelvin temperature},}\ }\href@noop {}
  {\bibfield  {journal} {\bibinfo  {journal} {Applied Physics Letters},\
  }\textbf {\bibinfo {volume} {100}},\ \bibinfo {pages} {243504} (\bibinfo
  {year} {2012})}\BibitemShut {NoStop}%
\bibitem [{\citenamefont {Tanner}(1976)}]{Tanner}%
  \BibitemOpen
  \bibfield  {author} {\bibinfo {author} {\bibfnamefont {B.}~\bibnamefont
  {Tanner}},\ }\href@noop {} {\emph {\bibinfo {title} {X-Ray Diffraction
  Topography}}}\ (\bibinfo  {publisher} {Pargamon},\ \bibinfo {address}
  {Oxford},\ \bibinfo {year} {1976})\BibitemShut {NoStop}%
\bibitem [{\citenamefont {Slavov}(1991)}]{slavov}%
  \BibitemOpen
  \bibfield  {author} {\bibinfo {author} {\bibfnamefont {S.}~\bibnamefont
  {Slavov}},\ }\bibfield  {title} {\enquote {\bibinfo {title} {Prediction of
  the frequency spectrum of at-cut contoured quartz resonators by means of
  x-ray diffraction topography},}\ }\href@noop {} {\bibfield  {journal}
  {\bibinfo  {journal} {Applied Physics A},\ }\textbf {\bibinfo {volume}
  {52}},\ \bibinfo {pages} {184} (\bibinfo {year} {1991})}\BibitemShut
  {NoStop}%
\bibitem [{\citenamefont {EerNisse}(1975)}]{1536996}%
  \BibitemOpen
  \bibfield  {author} {\bibinfo {author} {\bibfnamefont {E.~P.}\ \bibnamefont
  {EerNisse}},\ }\bibfield  {title} {\enquote {\bibinfo {title} {Quartz
  resonator frequency shifts arising from electrode stress},}\ }in\ \href@noop
  {} {\emph {\bibinfo {booktitle} {29th Annual Symposium on Frequency
  Control}}}\ (\bibinfo {year} {1975})\ pp.\ \bibinfo {pages} {1 --
  4}\BibitemShut {NoStop}%
\bibitem [{IEE(1987)}]{IEEE176}%
  \BibitemOpen
  \href@noop {} {\emph {\bibinfo {title} {IEEE Standard on Piezoelectricity,
  The Institute of Electrical and Electronics Engineers}}},\ \bibinfo
  {organization} {IEEE},\ \bibinfo {address} {345 East Street, New York, NY,
  10017, USA} (\bibinfo {year} {1987})\BibitemShut {NoStop}%
\bibitem [{\citenamefont {Besson}(1977)}]{1537081}%
  \BibitemOpen
  \bibfield  {author} {\bibinfo {author} {\bibfnamefont {R.~J.}\ \bibnamefont
  {Besson}},\ }\bibfield  {title} {\enquote {\bibinfo {title} {A new
  ``electrodeless" resonator design},}\ }in\ \Doi {10.1109/FREQ.1977.200141}
  {\emph {\bibinfo {booktitle} {31st Annual Symposium on Frequency Control}}}\
  (\bibinfo {year} {1977})\ pp.\ \bibinfo {pages} {147 -- 152}\BibitemShut
  {NoStop}%
\bibitem [{\citenamefont {Landau}\ and\ \citenamefont
  {Rumer}(1937)}]{landaurumer1}%
  \BibitemOpen
  \bibfield  {author} {\bibinfo {author} {\bibfnamefont {L.}~\bibnamefont
  {Landau}}\ and\ \bibinfo {author} {\bibfnamefont {G.}~\bibnamefont {Rumer}},\
  }\bibfield  {title} {\enquote {\bibinfo {title} {Uber schall absorption in
  festen {K}\"{o}rpen},}\ }\href@noop {} {\bibfield  {journal} {\bibinfo
  {journal} {Physikalische Zeitschrift der Sowjetunion},\ }\textbf {\bibinfo
  {volume} {11}},\ \bibinfo {pages} {18} (\bibinfo {year} {1937})}\BibitemShut
  {NoStop}%
\bibitem [{\citenamefont {Klemens}(1965)}]{landaurumer2}%
  \BibitemOpen
  \bibfield  {author} {\bibinfo {author} {\bibfnamefont {P.~G.}\ \bibnamefont
  {Klemens}},\ }\enquote {\bibinfo {title} {Physical acoustics},}\ \ (\bibinfo
  {publisher} {New York:Academic},\ \bibinfo {year} {1965})\ Chap.\ \bibinfo
  {chapter} {Effect of thermal and phonon processes on ultrasonic
  attennuation}\BibitemShut {NoStop}%
\bibitem [{\citenamefont {Maris}(1971)}]{Maris1971}%
  \BibitemOpen
  \bibfield  {author} {\bibinfo {author} {\bibfnamefont {H.}~\bibnamefont
  {Maris}},\ }\enquote {\bibinfo {title} {Physical acoustics},}\ \ (\bibinfo
  {publisher} {Academic},\ \bibinfo {year} {1971})\ Chap.\ \bibinfo {chapter}
  {Interaction of sound waves with thermal phonons in dielectric crystals},
  pp.\ \bibinfo {pages} {279--345}\BibitemShut {NoStop}%
\bibitem [{\citenamefont {Jiang}\ \emph {et~al.}(2004)\citenamefont {Jiang},
  \citenamefont {Yu}, \citenamefont {Liu},\ and\ \citenamefont
  {Huang}}]{PhysRevLett.93.185501}%
  \BibitemOpen
  \bibfield  {author} {\bibinfo {author} {\bibfnamefont {H.}~\bibnamefont
  {Jiang}}, \bibinfo {author} {\bibfnamefont {M.~F.}\ \bibnamefont {Yu}},
  \bibinfo {author} {\bibfnamefont {B.}~\bibnamefont {Liu}}, \ and\ \bibinfo
  {author} {\bibfnamefont {Y.}~\bibnamefont {Huang}},\ }\bibfield  {title}
  {\enquote {\bibinfo {title} {Intrinsic energy loss mechanisms in a
  cantilevered carbon nanotube beam oscillator},}\ }\href@noop {} {\bibfield
  {journal} {\bibinfo  {journal} {Physical Review Letters},\ }\textbf {\bibinfo
  {volume} {93}},\ \bibinfo {pages} {185501} (\bibinfo {year}
  {2004})}\BibitemShut {NoStop}%
\bibitem [{\citenamefont {H{\"u}ttel}\ \emph {et~al.}(2009)\citenamefont
  {H{\"u}ttel}, \citenamefont {Steele}, \citenamefont {Witkamp}, \citenamefont
  {Poot}, \citenamefont {Kouwenhoven},\ and\ \citenamefont {van~der
  Zant}}]{doi:10.1021/nl900612h}%
  \BibitemOpen
  \bibfield  {author} {\bibinfo {author} {\bibfnamefont {A.~K.}\ \bibnamefont
  {H{\"u}ttel}}, \bibinfo {author} {\bibfnamefont {G.~A.}\ \bibnamefont
  {Steele}}, \bibinfo {author} {\bibfnamefont {B.}~\bibnamefont {Witkamp}},
  \bibinfo {author} {\bibfnamefont {M.}~\bibnamefont {Poot}}, \bibinfo {author}
  {\bibfnamefont {L.~P.}\ \bibnamefont {Kouwenhoven}}, \ and\ \bibinfo {author}
  {\bibfnamefont {H.~S.~J.}\ \bibnamefont {van~der Zant}},\ }\bibfield  {title}
  {\enquote {\bibinfo {title} {Carbon nanotubes as ultrahigh quality factor
  mechanical resonators},}\ }\href@noop {} {\bibfield  {journal} {\bibinfo
  {journal} {Nano Letters},\ }\textbf {\bibinfo {volume} {9}},\ \bibinfo
  {pages} {2547} (\bibinfo {year} {2009})}\BibitemShut {NoStop}%
\bibitem [{\citenamefont {Shim}\ \emph {et~al.}(2007)\citenamefont {Shim},
  \citenamefont {Chun}, \citenamefont {Kang}, \citenamefont {Cho},
  \citenamefont {Park}, \citenamefont {Mohanty}, \citenamefont {Kim},\ and\
  \citenamefont {Kim}}]{shim:133505}%
  \BibitemOpen
  \bibfield  {author} {\bibinfo {author} {\bibfnamefont {S.~B.}\ \bibnamefont
  {Shim}}, \bibinfo {author} {\bibfnamefont {J.~S.}\ \bibnamefont {Chun}},
  \bibinfo {author} {\bibfnamefont {S.~W.}\ \bibnamefont {Kang}}, \bibinfo
  {author} {\bibfnamefont {S.~W.}\ \bibnamefont {Cho}}, \bibinfo {author}
  {\bibfnamefont {Y.~D.}\ \bibnamefont {Park}}, \bibinfo {author}
  {\bibfnamefont {P.}~\bibnamefont {Mohanty}}, \bibinfo {author} {\bibfnamefont
  {N.}~\bibnamefont {Kim}}, \ and\ \bibinfo {author} {\bibfnamefont
  {J.}~\bibnamefont {Kim}},\ }\bibfield  {title} {\enquote {\bibinfo {title}
  {Micromechanical resonators fabricated from lattice-matched and
  etch-selective gaas/ingap/gaas heterostructures},}\ }\href@noop {} {\bibfield
   {journal} {\bibinfo  {journal} {Applied Physics Letters},\ }\textbf
  {\bibinfo {volume} {91}},\ \bibinfo {pages} {133505} (\bibinfo {year}
  {2007})}\BibitemShut {NoStop}%
\bibitem [{\citenamefont {Zolfagharkhani}\ \emph {et~al.}(2005)\citenamefont
  {Zolfagharkhani}, \citenamefont {Gaidarzhy}, \citenamefont {Shim},
  \citenamefont {Badzey},\ and\ \citenamefont {Mohanty}}]{PhysRevB.72.224101}%
  \BibitemOpen
  \bibfield  {author} {\bibinfo {author} {\bibfnamefont {G.}~\bibnamefont
  {Zolfagharkhani}}, \bibinfo {author} {\bibfnamefont {A.}~\bibnamefont
  {Gaidarzhy}}, \bibinfo {author} {\bibfnamefont {S.~B.}\ \bibnamefont {Shim}},
  \bibinfo {author} {\bibfnamefont {R.~L.}\ \bibnamefont {Badzey}}, \ and\
  \bibinfo {author} {\bibfnamefont {P.}~\bibnamefont {Mohanty}},\ }\bibfield
  {title} {\enquote {\bibinfo {title} {Quantum friction in nanomechanical
  oscillators at millikelvin temperatures},}\ }\href@noop {} {\bibfield
  {journal} {\bibinfo  {journal} {Physical Review B},\ }\textbf {\bibinfo
  {volume} {72}},\ \bibinfo {pages} {224101} (\bibinfo {year}
  {2005})}\BibitemShut {NoStop}%
\bibitem [{\citenamefont {Imboden}\ and\ \citenamefont
  {Mohanty}(2009)}]{PhysRevB.79.125424}%
  \BibitemOpen
  \bibfield  {author} {\bibinfo {author} {\bibfnamefont {M.}~\bibnamefont
  {Imboden}}\ and\ \bibinfo {author} {\bibfnamefont {P.}~\bibnamefont
  {Mohanty}},\ }\bibfield  {title} {\enquote {\bibinfo {title} {Evidence of
  universality in the dynamical response of micromechanical diamond resonators
  at millikelvin temperatures},}\ }\href@noop {} {\bibfield  {journal}
  {\bibinfo  {journal} {Physical Review B},\ }\textbf {\bibinfo {volume}
  {79}},\ \bibinfo {pages} {125424} (\bibinfo {year} {2009})}\BibitemShut
  {NoStop}%
\bibitem [{\citenamefont {Seoanez}\ \emph {et~al.}(2007)\citenamefont
  {Seoanez}, \citenamefont {Guinea},\ and\ \citenamefont {{Castro
  Neto}}}]{0295-5075-78-6-60002}%
  \BibitemOpen
  \bibfield  {author} {\bibinfo {author} {\bibfnamefont {C.}~\bibnamefont
  {Seoanez}}, \bibinfo {author} {\bibfnamefont {F.}~\bibnamefont {Guinea}}, \
  and\ \bibinfo {author} {\bibfnamefont {A.~H.}\ \bibnamefont {{Castro
  Neto}}},\ }\bibfield  {title} {\enquote {\bibinfo {title} {Dissipation due to
  two-level systems in nano-mechanical devices},}\ }\href@noop {} {\bibfield
  {journal} {\bibinfo  {journal} {Europhysics Letters},\ }\textbf {\bibinfo
  {volume} {78}},\ \bibinfo {pages} {60002} (\bibinfo {year}
  {2007})}\BibitemShut {NoStop}%
\bibitem [{\citenamefont {Martin}(1984)}]{Martin}%
  \BibitemOpen
  \bibfield  {author} {\bibinfo {author} {\bibfnamefont {J.}~\bibnamefont
  {Martin}},\ }\bibfield  {title} {\enquote {\bibinfo {title} {Aluminum-related
  acoustic loss in at-cut quartz crystals},}\ }in\ \href@noop {} {\emph
  {\bibinfo {booktitle} {Proc. 38th Ann. Freq. Control Symposium}}}\ (\bibinfo
  {year} {1984})\ pp.\ \bibinfo {pages} {16--21}\BibitemShut {NoStop}%
\bibitem [{\citenamefont {Mason}(1990)}]{MasonWP}%
  \BibitemOpen
  \bibfield  {author} {\bibinfo {author} {\bibfnamefont {W.}~\bibnamefont
  {Mason}},\ }\bibfield  {title} {\enquote {\bibinfo {title} {Effects of
  impurities and phonon processes on the ultrasonic attenuation of germanium,
  crystal quartz, and silicon},}\ }in\ \href@noop {} {\emph {\bibinfo
  {booktitle} {Physical Acoustics}}},\ Vol.\ \bibinfo {volume} {III}\ (\bibinfo
   {publisher} {Elsevier Science},\ \bibinfo {year} {1990})\ Chap.\ \bibinfo
  {chapter} {Lattice Dynamics}\BibitemShut {NoStop}%
\bibitem [{\citenamefont {Euler}\ \emph {et~al.}(1982)\citenamefont {Euler},
  \citenamefont {Lipson}, \citenamefont {Kahan},\ and\ \citenamefont
  {Armington}}]{Euler}%
  \BibitemOpen
  \bibfield  {author} {\bibinfo {author} {\bibfnamefont {F.}~\bibnamefont
  {Euler}}, \bibinfo {author} {\bibfnamefont {H.}~\bibnamefont {Lipson}},
  \bibinfo {author} {\bibfnamefont {A.}~\bibnamefont {Kahan}}, \ and\ \bibinfo
  {author} {\bibfnamefont {A.}~\bibnamefont {Armington}},\ }\bibfield  {title}
  {\enquote {\bibinfo {title} {Characterization of alkali impurities in
  quartz},}\ }in\ \href@noop {} {\emph {\bibinfo {booktitle} {Proc. 36th Ann.
  on Freq. Contr. Symp.}}}\ (\bibinfo {year} {1982})\ pp.\ \bibinfo {pages}
  {115--123}\BibitemShut {NoStop}%
\bibitem [{\citenamefont {Halliburton}\ \emph {et~al.}(1979)\citenamefont
  {Halliburton}, \citenamefont {Markes}, \citenamefont {Martin}, \citenamefont
  {Doherty}, \citenamefont {Kouvakalis}, \citenamefont {Sibley}, \citenamefont
  {Armington},\ and\ \citenamefont {Brown}}]{Halliburton}%
  \BibitemOpen
  \bibfield  {author} {\bibinfo {author} {\bibfnamefont {L.}~\bibnamefont
  {Halliburton}}, \bibinfo {author} {\bibfnamefont {M.}~\bibnamefont {Markes}},
  \bibinfo {author} {\bibfnamefont {J.}~\bibnamefont {Martin}}, \bibinfo
  {author} {\bibfnamefont {S.}~\bibnamefont {Doherty}}, \bibinfo {author}
  {\bibfnamefont {N.}~\bibnamefont {Kouvakalis}}, \bibinfo {author}
  {\bibfnamefont {W.}~\bibnamefont {Sibley}}, \bibinfo {author} {\bibfnamefont
  {A.}~\bibnamefont {Armington}}, \ and\ \bibinfo {author} {\bibfnamefont
  {R.}~\bibnamefont {Brown}},\ }\bibfield  {title} {\enquote {\bibinfo {title}
  {Radiation effects in synthetic quartz: The role of electrodiffusion and
  radiation-induced mobility of interstitial ions},}\ }\href@noop {} {\bibfield
   {journal} {\bibinfo  {journal} {IEEE Transactions on Nuclear Science},\
  }\textbf {\bibinfo {volume} {26}},\ \bibinfo {pages} {4851} (\bibinfo {year}
  {1979})}\BibitemShut {NoStop}%
\bibitem [{\citenamefont {Welton}(1973)}]{Welton1973}%
  \BibitemOpen
  \bibfield  {author} {\bibinfo {author} {\bibfnamefont {P.}~\bibnamefont
  {Welton}},\ }\bibfield  {title} {\enquote {\bibinfo {title} {The potential
  method formulation of acoustic wave scattering by rough surfaces},}\
  }\href@noop {} {\bibfield  {journal} {\bibinfo  {journal} {Journal of the
  Acoustical Society of America},\ }\textbf {\bibinfo {volume} {54}},\ \bibinfo
  {pages} {66} (\bibinfo {year} {1973})}\BibitemShut {NoStop}%
\bibitem [{\citenamefont {Deresiewicz}(1954)}]{Deresiewicz1954}%
  \BibitemOpen
  \bibfield  {author} {\bibinfo {author} {\bibfnamefont {H.}~\bibnamefont
  {Deresiewicz}},\ }\bibfield  {title} {\enquote {\bibinfo {title} {Plane waves
  in a thermoelastic solid},}\ }\href@noop {} {\bibfield  {journal} {\bibinfo
  {journal} {Journal of the Acoustical Society of America},\ }\textbf {\bibinfo
  {volume} {29}},\ \bibinfo {pages} {204} (\bibinfo {year} {1954})}\BibitemShut
  {NoStop}%
\bibitem [{\citenamefont {Barron}\ \emph {et~al.}(1982)\citenamefont {Barron},
  \citenamefont {J.F.Collins}, \citenamefont {T.W.Smith},\ and\ \citenamefont
  {G.K.White}}]{Barron1982}%
  \BibitemOpen
  \bibfield  {author} {\bibinfo {author} {\bibfnamefont {T.}~\bibnamefont
  {Barron}}, \bibinfo {author} {\bibnamefont {J.F.Collins}}, \bibinfo {author}
  {\bibnamefont {T.W.Smith}}, \ and\ \bibinfo {author} {\bibnamefont
  {G.K.White}},\ }\bibfield  {title} {\enquote {\bibinfo {title} {Thermal
  expansion, gr{\"u}neisen functions and static lattice properties of
  quartz},}\ }\href@noop {} {\bibfield  {journal} {\bibinfo  {journal} {Journal
  of Physics C: Solid State Physics},\ }\textbf {\bibinfo {volume} {15}},\
  \bibinfo {pages} {4311} (\bibinfo {year} {1982})}\BibitemShut {NoStop}%
\bibitem [{\citenamefont {Zeller}\ and\ \citenamefont
  {Pohl}(1971)}]{Zeller1971}%
  \BibitemOpen
  \bibfield  {author} {\bibinfo {author} {\bibfnamefont {R.}~\bibnamefont
  {Zeller}}\ and\ \bibinfo {author} {\bibfnamefont {R.}~\bibnamefont {Pohl}},\
  }\bibfield  {title} {\enquote {\bibinfo {title} {Thermal conductivity and
  specific heat of noncrystalline solids},}\ }\href@noop {} {\bibfield
  {journal} {\bibinfo  {journal} {Physical Review B},\ }\textbf {\bibinfo
  {volume} {4}},\ \bibinfo {pages} {2029} (\bibinfo {year} {1971})}\BibitemShut
  {NoStop}%
\end{thebibliography}

%


\listoffigures

\end{document}